\documentclass[12pt]{iopart}
\usepackage{iopams}
\begin{document}

\title{Black hole evaporation in a spherically symmetric 
non-commutative space-time}

\author{Elisabetta Di Grezia\dag, Giampiero Esposito\ddag, 
Gennaro Miele\ddag}

\address{\dag Universit\`a Statale di Bergamo, Facolt\`a di Ingegneria,
Viale Marconi 5, 24044 Dalmine (Bergamo); INFN, Sezione di Milano,
Via Celoria 16, 20133 Milano}

\address{\ddag INFN, Sezione di Napoli, and Dipartimento di 
Scienze Fisiche, Complesso Universitario di Monte
S. Angelo, Via Cintia, Edificio 6, 80126 Napoli, Italy}

\begin{abstract}
Recent work in the literature has studied the quantum-mechanical decay of
a Schwarzschild-like black hole, formed by gravitational collapse, into
almost-flat space-time and weak radiation at a very late time. The relevant
quantum amplitudes have been evaluated for bosonic and fermionic fields,
showing that no information is lost in collapse to a black hole. On the
other hand, recent developments in noncommutative geometry have shown
that, in general relativity, the effects of noncommutativity can be taken
into account by keeping the standard form of the Einstein tensor on the
left-hand side of the field equations and introducing a modified
energy-momentum tensor as a source on the right-hand side. 
Relying on the recently obtained noncommutativity effect on a
static, spherically symmetric metric, 
we have considered from a new perspective the
quantum amplitudes in black hole evaporation. The general relativity
analysis of spin-2 amplitudes has been 
shown to be modified by a multiplicative
factor $F$ depending on a constant non-commutativity parameter and on the
upper limit $R$ of the radial coordinate. Limiting forms of $F$ have been
derived which are compatible with the adiabatic approximation.
\end{abstract}

\section{Introduction}

Theoretical research in black hole physics has witnessed,
over the last few years, an impressive
amount of new ideas and results on at least four main areas:
\vskip 0.3cm
\noindent
(i) The problem of information loss in black holes, after the suggestion
in Ref. \cite{Hawk05} that quantum gravity is unitary and information is
preserved in black hole formation and evaporation.
\vskip 0.3cm
\noindent
(ii) The related series of papers in Refs. \cite{Farl1, Farl2, Farl3, Farl4,
Farl5, Farl6, Farl7, Farl8, Farl9}, concerned with evaluating quantum
amplitudes for transitions from initial to final states, in agreement with
a picture where information is not lost, and the end state of black hole
evaporation is a combination of outgoing radiation states.
\vskip 0.3cm
\noindent
(iii) The approach in Refs. \cite{Vilk1, Vilk2, Vilk3}, according to which
black holes create instead a vacuum matter charge to protect themselves
from the quantum evaporation.
\vskip 0.3cm
\noindent
(iv) The work in Ref. \cite{Nico06} where the authors, relying upon the
previous findings in Ref. \cite{Smai04}, consider a non-commutative radiating
Schwarzschild black hole, and find that non-commutativity cures usual problems
encountered in trying to describe the latest stage of black hole evaporation.

We have been therefore led to study how non-commutativity would affect the
analysis of quantum amplitudes in black hole evaporation performed in
Refs. \cite{Farl1, Farl2, Farl3, Farl4, Farl5, Farl6, Farl7, Farl8, Farl9}.
Following Ref. \cite{Nico06}, we assume that non-commutativity of space-time
can be encoded in the commutator of operators corresponding to space-time
coordinates, i.e. (the integer $D$ below is even)
\begin{equation}
\Bigr[x^{\mu},x^{\nu}\Bigr]=i \theta^{\mu \nu}, \;
\mu,\nu=1,2,...,D,
\label{(1)}
\end{equation}
where the antisymmetric matrix $\theta^{\mu \nu}$ is taken to have
block-diagonal form
\begin{equation}
\theta^{\mu \nu}={\rm diag}\Bigr(\theta_{1},...,\theta_{D/2}\Bigr),
\label{(2)}
\end{equation}
with
\begin{equation}
\theta_{i}=\theta
\left(
\begin{array}{cc}
0 & 1 \\
-1 & 0 \\
\end{array}
\right), \;
\forall i =1,2,...,D/2,
\label{(3)}
\end{equation}
the parameter $\theta$ having dimension of length squared and being
constant. As shown in Ref. \cite{Smai04}, the constancy of $\theta$ leads
to a consistent treatment of Lorentz invariance and unitarity.
The authors of Ref. \cite{Nico06} solve the Einstein equations with mass
density of a static, spherically symmetric, smeared particle-like
gravitational source as (hereafter we work in $G=c={\hbar}=1$ units)
\begin{equation}
\rho_{\theta}(r)={M\over (4\pi \theta)^{3\over 2}}e^{-{r^{2}\over 4\theta}},
\label{(4)}
\end{equation}
which therefore plays the role of matter source. Their resulting
spherically symmetric metric is
\begin{eqnarray}
ds^{2}&=&-\left[1-{4M\over r \sqrt{\pi}}\gamma
\left({3\over 2},{r^{2}\over 4\theta}\right)\right]dt^{2}
+\left[1-{4M\over r \sqrt{\pi}}\gamma \left({3\over 2},
{r^{2}\over 4 \theta}\right)\right]^{-1}dr^{2} \nonumber \\
&+& r^{2}(d\Theta^{2}+\sin^{2}\Theta d\phi^{2}),
\label{(5)}
\end{eqnarray}
where we use the lower incomplete gamma function \cite{Nico06}
\begin{equation}
\gamma \left({3\over 2},{r^{2}\over 4\theta}\right) \equiv
\int_{0}^{{r^{2}\over 4\theta}}\sqrt{t}e^{-t}dt.
\label{(6)}
\end{equation}
In this picture, one deals with a mass distribution
\begin{equation}
m(r) \equiv {2M \over \sqrt{\pi}}\gamma \left({3\over 2},
{r^{2}\over 4\theta}\right),
\label{(7)}
\end{equation}
while $M$ is the total mass of the source \cite{Nico06}. This mass
function satisfies the equation
$$
m'(r)=4\pi r^{2} \rho_{\theta}(r),
$$
formally analogous to the general relativity case \cite{Farl9}.

The work in Refs. \cite{Farl1, Farl2, Farl3, Farl4, Farl5, Farl6, Farl7,
Farl8, Farl9} studies instead the quantum-mechanical decay of a
Schwarzschild-like black hole, formed by gravitational collapse, into
almost-flat space-time and weak radiation at a very late time.
The spin-2 gravitational perturbations split into parts with odd and
even parity, and one can isolate suitable variables which can be taken
as boundary data on a final spacelike hypersurface $\Sigma_{F}$. The main
idea is then to consider a complexified classical boundary-value problem
where $T$ is rotated into the complex: $T \rightarrow |T| e^{-i \alpha}$,
for $\alpha \in ]0,\pi/2]$, and evaluate the corresponding classical
Lorentzian action $S_{\rm class}^{(2)}$
to quadratic order in metric perturbations. The
genuinely Lorentzian quantum amplitude is recovered by taking the limit
as $\alpha \rightarrow 0^{+}$ of the semiclassical amplitude
$e^{iS_{\rm class}^{(2)}}$ \cite{Farl1, Farl5, Farl9}.

Section II studies the differential equations obeyed by radial modes
within the framework of the adiabatic approximation, and Sec. III obtains
the resulting orthogonality relation in the presence of a non-vanishing
non-commutativity parameter $\theta$. Section IV derives the effect of
$\theta$ on the expansion of the pure-gravity action functional, which
can be used in the evaluation of quantum amplitudes along the lines of
Refs. \cite{Farl1}-\cite{Farl9}, while
concluding remarks are presented in Sec. V.

\section{Equations for radial modes}

The analysis in Ref. \cite{Farl9} holds for any
spherically symmetric Lorentzian background metric
\begin{equation}
ds^{2}=-e^{b(r,t)}dt^{2}+e^{a(r,t)}dr^{2}
+r^{2}(d\Theta^{2}+\sin^{2}\Theta d\phi^{2}),
\label{(8)}
\end{equation}
the even modes $\xi_{2lm}^{(+)}(r,t)$ and odd modes $\xi_{2lm}^{(-)}(r,t)$
being built from a Fourier-type decomposition, i.e. \cite{Farl9}
\begin{equation}
\xi_{2lm}^{(+)}(r,t)=\int_{-\infty}^{\infty}dk \; a_{2klm}^{(+)}
\xi_{2kl}^{(+)}(r){\sin kt \over \sin kT},
\label{(9)}
\end{equation}
and
\begin{equation}
\xi_{2lm}^{(-)}(r,t)=\int_{-\infty}^{\infty}dk \;
a_{2klm}^{(-)}\xi_{2kl}^{(-)}(r){\cos kt \over \sin kT},
\label{(10)}
\end{equation}
where the radial functions $\xi_{2kl}^{(\pm)}$ obey the following
second-order differential equation:
\begin{equation}
e^{-a}{d\over dr}\left(e^{-a}{d\xi_{2kl}^{(\pm)}\over dr}\right)
+\Bigr(k^{2}-V_{l}^{\pm}(r)\Bigr)\xi_{2kl}^{(\pm)}=0,
\label{(11)}
\end{equation}
where, on defining $\lambda \equiv {(l+2)(l-1)\over 2}$, the potential
terms are given by \cite{Farl9}
\begin{equation}
V_{l}^{+}(r)=e^{-a(r,t)}
{2[\lambda^{2}(\lambda+1)r^{3}+3\lambda^{2}mr^{2}+9]m^{2}r+9m^{3}
\over r^{3}(\lambda r+3m)^{2}},
\label{(12)}
\end{equation}
and
\begin{equation}
V_{l}^{-}(r)=e^{-a(r,t)}\left({l(l+1)\over r^{2}}
-{6m\over r^{3}}\right),
\label{(13)}
\end{equation}
respectively. In the expansion of the gravitational action to quadratic
order, it is of crucial importance to evaluate the integral
\begin{equation}
I(k,k',l,R) \equiv \int_{0}^{R}
e^{a(r,t)}\xi_{2kl}^{(+)}(r)\xi_{2k'l}^{(+)}(r)dr,
\label{(14)}
\end{equation}
since \cite{Farl9} 
\begin{equation}
S_{\rm class}^{(2)}[(h_{ij}^{(\pm)})_{lm}]
={\pm 1 \over 32 \pi} \sum_{l=2}^{\infty}\sum_{m=-l}^{l}
{(l-2)! \over (l+2)!}\int_{0}^{R}e^{a}\xi_{2lm}^{(\pm)}
\left({\partial \over \partial t}\xi_{2lm}^{(\pm)*}\right)_{t=T}dr.
\label{(15)}
\end{equation}
For this purpose, we bear in mind the limiting behaviours \cite{Farl9}
\begin{equation}
\xi_{2kl}^{(\pm)} \sim {\rm const} \times (kr)^{l+1}
+{\rm O}((kr)^{l+3}) \; {\rm as} \; r \rightarrow 0,
\label{(16)}
\end{equation}
\begin{equation}
\xi_{2kl}^{(\pm)}(r) \sim z_{2kl}^{(\pm)}e^{ik r_{s}}
+z_{2kl}^{(\pm)*}e^{-ikr_{s}} \; {\rm as} \; r \rightarrow \infty,
\label{(17)}
\end{equation}
where Eq. (16) results from imposing regularity at the origin,
$r_{s}$ is the Regge--Wheeler tortoise coordinate \cite{Farl9, Regg57}
\begin{equation}
r_{s}(r) \equiv r +2M \log(r-2M),
\label{(18)}
\end{equation}
while $z_{2kl}^{(\pm)}$ are complex constants. Indeed, it should be
stressed that non-commutativity can smear plane waves into Gaussian wave
packets. Thus, in a fully self-consistent analysis, the Fourier modes in
Eq. (9), (10), and their asymptotic form in Eq. (16), (17), should be
modified accordingly. However, this task goes beyond the aims of the
present paper, and is deferred to a future publication.

With this understanding, we can now exploit Eq. (11)
to write the equations (hereafter, we write for simplicity of notation
$\xi_{kl}$ rather than $\xi_{2kl}^{(\pm)}$, and similarly for $V_{l}$
rather than $V_{l}^{\pm}$)
\begin{equation}
e^{a}\xi_{k'l}\left[e^{-a}{d\over dr}\left(e^{-a}{d\over dr}\xi_{kl}\right)
+(k^{2}-V_{l})\xi_{kl}\right]=0,
\label{(19)}
\end{equation}
\begin{equation}
e^{a}\xi_{kl}\left[e^{-a}{d\over dr}\left(e^{-a}{d\over dr}\xi_{k'l}\right)
+({k'}^{2}-V_{l})\xi_{k'l}\right]=0.
\label{(20)}
\end{equation}
According to a standard procedure, if we subtract Eq. (20) from Eq. (19),
and integrate the resulting equation from $r=0$ to $r=R$, we obtain
\begin{equation}
(k^{2}-{k'}^{2})\int_{0}^{R}e^{a}\xi_{kl}\xi_{k'l}dr
= \int_{0}^{R}\left[\xi_{kl}{d\over dr}\left(e^{-a}{d\over dr}
\xi_{k'l}\right)-\xi_{k'l}{d\over dr}\left(e^{-a}{d\over dr}\xi_{kl}
\right)\right]dr.
\label{(21)}
\end{equation}
The desired integral (14) is therefore obtained from Eq. (21),
whose right-hand side is then
completely determined from the limiting behaviours in
Eqs. (16) and (17), i.e.
\begin{equation}
\int_{0}^{R}e^{a}\xi_{kl}\xi_{k'l}dr=
\left \{ {1\over (k^{2}-{k'}^{2})}\left[\xi_{kl}e^{-a}
\left({d\over dr}\xi_{k'l}\right)-\xi_{k'l}e^{-a}
\left({d\over dr}\xi_{kl}\right)\right]_{r=0}^{r=R}
\right \},
\label{(22)}
\end{equation}
where, on going from Eq. (21) to Eq. (22), we have exploited the
vanishing coefficient that weights the integral
$$
\int_{0}^{R}e^{-a}\left({d\over dr}\xi_{kl}\right)
\left({d\over dr}\xi_{k'l}\right)dr,
$$
resulting from two contributions of equal magnitude and opposite sign.
By virtue of Eq. (16), $r=0$ gives vanishing contribution to the right-hand
side of Eq. (22), while the contribution of first derivatives of radial
functions involves also
\begin{equation}
\left . {dr_{s}\over dr} \right |_{r=R}
={R\over (R-2M)}.
\label{(23)}
\end{equation}

\section{Generalized orthogonality relation}

Note now that our metric (5) is a particular case of the spherically
symmetric metric (8), since our $a$ and $b$ functions are independent
of time. More precisely, unlike the full Vaidya space-time, where in the
region containing outgoing radiation the mass function varies extremely
slowly with respect both to $t$ and to $r$
\cite{Farl8}, we consider a ``hybrid''
scheme where the mass function depends on $r$ only for any fixed value of
the non-commutativity parameter $\theta$. We can therefore write
\begin{equation}
e^{-a}=1-{4M\over r \sqrt{\pi}}
\gamma \left({3\over 2},{r^{2}\over 4\theta}\right)
\label{(24)}
\end{equation}
in our non-commutative spherically symmetric model,
where the function in curly brackets in Eq. (22) reads as
$$
\left(1-{4M \over R \sqrt{\pi}}\gamma \left({3\over 2},
{R^{2}\over 4\theta}\right)\right){R \over (R-2M)}
{1\over (k^{2}-{k'}^{2})}
\times i \Bigr[(k'-k)z_{kl}z_{k'l}e^{i(k+k')r_{s}(R)}
$$
$$
+(k-k')z_{kl}^{*}z_{k'l}^{*}e^{-i(k+k')r_{s}(R)}
-(k+k')z_{kl}z_{k'l}^{*}e^{i(k-k')r_{s}(R)}
+(k+k')z_{kl}^{*}z_{k'l}e^{i(k'-k)r_{s}(R)}\Bigr].
$$
At this stage, we exploit one of the familiar limits that can be used
to express the Dirac $\delta$, i.e. \cite{Farl9}
\begin{equation}
\lim_{r_{s}\to \infty}{e^{i(k \pm k')r_{s}}\over (k \pm k')}
=i \pi \delta(k \pm k'),
\label{(25)}
\end{equation}
to find
\begin{equation}
\int_{0}^{R}e^{a}\xi_{kl}\xi_{k'l}dr=2\pi |z_{kl}|^{2}
F(R,\theta) \Bigr(\delta(k+k')+\delta(k-k')\Bigr),
\label{(26)}
\end{equation}
having defined
\begin{equation}
F(R,\theta) \equiv {R \over (R-2M)} \left[1-
{4M \over R \sqrt{\pi}}\gamma \left({3\over 2},{R^{2}\over 4\theta}\right)
\right].
\label{(27)}
\end{equation}

\section{Effect of $\theta$ and expansion of the action functional}

Since $\theta$ has dimension length squared as we said after Eq. (3),
we can define the non-commutativity-induced length scale
\begin{equation}
L \equiv 2 \sqrt{\theta}.
\label{(28)}
\end{equation}
Moreover, we know that
our results only hold in the adiabatic approximation, i.e. when both
$m'$ and ${\dot m}$ are very small. The latter condition is obviously
satisfied because our mass function in Eq. (7) is independent of time.
The former amounts to requiring that (hereafter we set
$w \equiv R/L$, while $R_{s} \equiv 2M$)
\begin{equation}
m'(R)={2\over \sqrt{\pi}}{R_{s}\over L}e^{-w^{2}}w^{2} <<1.
\label{(29)}
\end{equation}
The condition (29) is satisfied provided that either
\vskip 0.3cm
\noindent
(i) $w \rightarrow \infty$ or $w \rightarrow 0$, i.e. $R >>L$ or
$R << L$;
\vskip 0.3cm
\noindent
(ii) or at $R=L$ such that
\begin{equation}
m'(R=L)=m'(w=1)={2\over \sqrt{\pi}}{R_{s}\over L}e^{-1} <<1,
\label{(30)}
\end{equation}
and hence for ${R_{s}\over L} << {e \sqrt{\pi} \over 2}$.

Furthermore, at finite values of the non-commutativity parameter $\theta$,
our $w \equiv {R\over L}$ is always much larger than $1$ in Eq. (27)
if $R$ is very large, and hence
we can exploit the asymptotic expansion of the lower incomplete
$\gamma$-function in this limit \cite{Abra64, Spal06}, i.e.
\begin{equation}
\gamma \left({3\over 2},w^{2}\right)=
\Gamma \left ({3\over 2} \right)
-\Gamma \left({3\over 2},w^{2}\right)
\sim {1\over 2}\sqrt{\pi}\left[1-e^{-w^{2}}
\sum_{p=0}^{\infty}{w^{1-2p}
\over \Gamma \left({3\over 2}-p \right)}\right].
\label{(31)}
\end{equation}
By virtue of Eqs. (27) and (31), we find
\begin{equation}
F(R,\theta) \equiv F(R,L) \sim 1+{R_{s} \over (R-R_{s})}
e^{-w^{2}}
\sum_{p=0}^{\infty}{w^{1-2p}\over
\Gamma \left({3\over 2}-p \right)}.
\label{(32)}
\end{equation}
Equation (32) describes the asymptotic expansion
of the correction factor $F$ when $R >> L$.

In the opposite regime, i.e. for $\theta$ so large that
$(R/L)<<1$ despite that $R$ tends to $\infty$, one has
\cite{Spal06}
\begin{equation}
F(R,L) \sim {R \over (R-R_{s})}\left[
1-{4 \over 3 \sqrt{\pi}}{R_{s}\over R}w^{3}
\left(1-{7\over 5}w^{2}\right)\right].
\label{(33)}
\end{equation}

Last, but not least, if $R$ and $L$ are comparable, the lower-incomplete
$\gamma$-function in Eq. (27) cannot be expanded, and we find,
bearing in mind that $R_{s}/L <<1$ from Eq. (30), the limiting form
\begin{equation}
F(R,L) \sim 1+ {R_{s}\over L} \left(1-{2\over \sqrt{\pi}}
\gamma \left({3\over 2},1 \right)\right)
+{\rm O}((R_{s}/L)^{2}).
\label{(34)}
\end{equation}

We therefore conclude that a $\theta$-dependent correction to the
general relativity analysis in Ref. \cite{Farl9} does indeed arise from
non-commutative geometry. In particular, the expansion of the
action to quadratic order in perturbative modes takes the form
(cf. Ref. \cite{Farl9})
\begin{eqnarray}
S_{\rm class}^{(2)}&=&{F(R,L)\over 16} \sum_{l=2}^{\infty}
\sum_{m=-l}^{l}\sum_{P=\pm 1}{(l-2)!\over (l+2)!} \times \nonumber \\
& \times & \int_{0}^{\infty}dk \;
k |z_{2klP}|^{2} \left | a_{2klmP}+P a_{2,-klmP} \right |^{2} \cot kT,
\label{(35)}
\end{eqnarray}
where the function $F(R,L)$ (see Eq. (27)) takes the
limiting forms (32) and (33), respectively, depending on whether
$w>>1$ or $w<<1$, while
$P = \pm 1$ for even (respectively odd) metric perturbations. Our
``correction'' $F(R,L)$ to the general relativity analysis is
non-vanishing provided that one works at very large but finite values
of $R$. In the limit as $R \rightarrow \infty$, one has instead
\begin{equation}
\lim_{R \to \infty} F(R,L)=1,
\label{(36)}
\end{equation}
which means that, at infinite distance from the Lorentzian singularity
of Schwarzschild geometry, one cannot detect the effect of a
non-commutativity parameter.

\section{Concluding remarks}

Our paper has investigated the effect of non-commutative geometry on the
recent theoretical analysis of quantum amplitudes in black hole
evaporation, following the work in Refs. \cite{Hawk05},
\cite{Farl1}-\cite{Farl9} (for other developments, see for example the
recent work in Refs. \cite{Kar05, Kar06a, Kar06b, Albe06}).
For this purpose, we have considered an
approximate scheme where the background space-time is static and
spherically symmetric, with mass function depending on the radial
coordinate only for any fixed value of the non-commutativity parameter
$\theta$.

Within this framework, we find \cite{DiGr06} 
that the general relativity analysis
of spin-2 amplitudes is modified by a multiplicative factor $F$ defined
in Eq. (27). Its limiting forms for $R >>L$ or $R <<L$ or
$R \cong L$ are given by Eqs. (32), (33) and (34), respectively.
Within this framework, {\it unitarity is preserved, and the end state of
black hole evaporation is a combination of outgoing radiation states}
(see section 1).

An outstanding open problem is
whether one can derive a time-dependent spherically
symmetric background metric which incorporates the effects of
non-commutative geometry. This would make it possible to improve the
present comparison with the results in Refs. \cite{Farl1}-\cite{Farl9},
where the Vaidya space-time was taken as the background geometry.
More recently, for the scalar wave equation in a non-commutative
spherically symmetric space-time, we have built the associated
conformal infinity \cite{DiGr07}, and the analysis of the wave equation
has been reduced to the task of solving an inhomogeneous 
Euler--Poisson--Darboux equation. The scalar field has been found to have
an asymptotic behaviour with a fall-off going on rather more slowly than
in flat space-time, in full qualitative agreement with general relativity
\cite{Schm79}.

\section*{Acknowledgments}
The work of G. Miele has been partially supported by
PRIN {\it FISICA ASTROPARTICELLARE}.


\section*{References}

\begin{thebibliography}{99}
\bibitem{Hawk05}
Hawking S W 2005 {\it Phys. Rev.} D {\bf 72} 084013
\bibitem{Farl1}
Farley A N St J and D'Eath P D 2004 {\it Phys. Lett.} B {\bf 601} 184
\bibitem{Farl2}
Farley A N St J and D'Eath P D 2005 {\it Phys. Lett.} B {\bf 613} 181
\bibitem{Farl3}
Farley A N St J and D'Eath P D 2005 {\it Class. Quantum Grav.} {\bf 22}
2765
\bibitem{Farl4}
Farley A N St J and D'Eath P D 2005 {\it Class. Quantum Grav.} {\bf 22}
3001
\bibitem{Farl5}
Farley A N St J and D'Eath P D 2005 `Quantum amplitudes in black-hole
evaporation. I. Complex approach' (gr-qc/0510028)
\bibitem{Farl6}
Farley A N St J and D'Eath P D 2005 `Quantum amplitudes in black-hole
evaporation. II. Spin-0 amplitude' (gr-qc/0510029)
\bibitem{Farl7}
Farley A N St J and D'Eath P D 2006 {\it Phys. Lett.} B {\bf 634} 419
\bibitem{Farl8}
Farley A N St J and D'Eath P D 2006 {\it Gen. Rel. Grav.} {\bf 38} 425
\bibitem{Farl9}
Farley A N St J and D'Eath P D 2006 {\it Ann. Phys. (N.Y.)} {\bf 321}
1334
\bibitem{Vilk1}
Vilkovisky G A 2006 {\it Ann. Phys. (N.Y.)} {\bf 321} 2717 
\bibitem{Vilk2}
Vilkovisky G A 2006 {\it Phys. Lett.} B {\bf 634} 456
\bibitem{Vilk3}
Vilkovisky G A 2006 Phys. Lett. B {\bf 638} 523
\bibitem{Nico06}
Nicolini P, Smailagic A, and Spallucci E 2006 {\it Phys. Lett.} B {\bf 632}
547
\bibitem{Smai04}
Smailagic A and Spallucci E 2004 {\it J. Phys.} A {\bf 37} 7169
\bibitem{Regg57}
Regge T and Wheeler J A 1957 {\it Phys. Rev.} {\bf 108} 1063
\bibitem{Abra64}
Abramowitz M and Stegun I A 1964 {\it Handbook of Mathematical Functions}
(New York: Dover)
\bibitem{Spal06}
Spallucci E, Smailagic A and Nicolini P 2006 {\it Phys. Rev.} D {\bf 73}
084004
\bibitem{Kar05}
Kar S and Majumdar S 2006 {\it Int. J. Mod. Phys.} A {\bf 21} 6087 
\bibitem{Kar06a}
Kar S and Majumdar S 2006 {\it Phys. Rev.} D {\bf 74} 066003
\bibitem{Kar06b}
Kar S 2006 {\it Phys. Rev.} D {\bf 74} 126002
\bibitem{Albe06}
Alberghi G L, Casadio R, Galli D, Gregori D, Tronconi A, and Vignoni V
2006 `Probing quantum gravity effects in black holes at LHC'
(hep-ph/0601243).
\bibitem{DiGr06}
Di Grezia E, Esposito G, and Miele G 2006 {\it Class. Quantum Grav.} 
{\bf 23} 6425
\bibitem{DiGr07}
Di Grezia E, Esposito G, and Miele G 2007 arXiv:0705.0242 [hep-th]
\bibitem{Schm79}
Schmidt B G and Stewart J M 1979 {\it Proc. Roy. Soc. Lond.} A
{\bf 367} 503
\end{thebibliography}
\end{document}